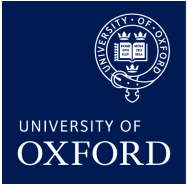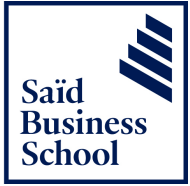

# The Oxford Olympics Study 2024: Are Cost and Cost Overrun at the Games Coming Down?

July 2024

Alexander Budzier, University of Oxford

Bent Flyvbjerg, University of Oxford, IT University of Copenhagen

# Abstract

The present paper is an update and extension of the "The Oxford Olympics Study 2016" (Flyvbjerg et al. 2016). We document that the Games remain costly and continue to have large cost overruns, to a degree that threatens their viability. The IOC is aware of the problem and has initiated reform. We assess the reforms and find: (a) Olympic costs are statistically significantly increasing; prior analyses did not show this trend; it is a step in the wrong direction. (b) Cost overruns were decreasing until 2008, but have increased since then; again a step in the wrong direction. (c) At present, the cost of Paris 2024 is USD 8.7 billion (2022 level) and cost overrun is 115% in real terms; this is not the frugal Games that were promised. (d) Cost overruns are the norm for the Olympics, past, present, and future; the Games are the only project type that never delivered on budget, ever. We assess a new IOC policy of reducing cost by reusing existing venues instead of building new ones. We find that reuse did not have the desired effect for Tokyo 2020 and also looks ineffective for Paris 2024. Finally, we recommend that the Games look to other types of megaproject for better data, better forecasting, and better methods for generating the positive learning curves that are necessary for bringing costs and overrun down. Only if this happens are Los Angeles 2028 and Brisbane 2032 likely to live up to the IOC's intentions of a more affordable Games that more cities will want to host.

# Why Study Cost and Cost Overrun at the Olympics?

In July 2023, Victoria, Australia surprisingly cancelled the 2026 Commonwealth Games. The cost had reportedly increased from AUD 2.6 billion to AUD 6-7 billion for the 12-day event (Kelly et al. 2023). Less than a year later, in March 2024, media reports emerged that Brisbane considered pulling out of the 2032 Summer Olympics due to rising costs. The Queensland Government and the International Olympic Committee quickly refuted the news (Kirk 2024).

Our previous studies (Flyvbjerg et al. 2021, 2016; Flyvbjerg and Stewart 2012) have documented high cost and cost overrun at the Olympic Games. This was initially a contested finding. Today, it is broadly accepted in the academic debate (see Appendix A). High cost and cost overrun have been argued to threaten the economic sustainability of the Games (Müller et al. 2021).

Given that the last three Summer Games cost USD 51 billion (in 2022 prices) and overran budgets by 185% in real terms – not including road, rail, airport, hotel, and other infrastructure, which often cost more than the Games themselves – the financial size and risks of the Games warrant study.

Understanding the magnitude of cost overruns in relation to the original budget is crucial to assessing financial risk for prospective host cities. The Paris 2024 Games, for instance, have seen costs surge from EUR 3.6 billion to 8.8 billion. Similarly, Los Angeles 2028 has revised its forecast from USD 5.3 billion to 6.8 billion.[1] These figures underscore the risks and challenges that host cities and national governments must foresee and manage.

As part of bidding to host the Games, the International Olympic Committee (IOC) requires host cities and governments to guarantee that they will cover any overruns to the Olympic budgets. The Olympic Host Contract locks the hosts into a non-negotiable commitment to cover any such increases. If overruns are likely, hosts and the IOC must consider cost escalation in planning the Games to get a realistic picture of the final outturn costs. Hosts typically set aside a contingency of 10%-15%. Historically, this has proven inadequate. Presently, cost escalation on the 2024, 2028, and 2032 Summer Games are already above this level of contingency.

Moreover, Flyvbjerg et al. (2016) argued that given the current global economic climate and subsequent tightening of government spending in many countries, understanding the implications of significant investments like the Games is critical for governments to make sound fiscal and economic decisions about their expenditures. The global economic climate has not changed since 2016, in fact, COVID-19 has only further restricted governments' ability to spend (World Bank 2022). For instance, cost overrun and associated debt from the Athens 2004 Games weakened the Greek economy and contributed to the country's deep financial and economic crises, beginning in 2007 and still playing out almost a decade later (Flyvbjerg 2011). For Rio 2016, the Brazilian economy was doing well when the city bid for the Olympics. Fast forward a decade to two months before the opening ceremony and this was no longer the case. Rio was now in such dire straits that the governor declared a state of emergency to secure additional funding for the Games from money reserved for dealing with natural and other disasters (Zimbalist 2020). Other hosts – especially those in small and weak economies – may want to ensure they do not end up in a similar situation by having a realistic picture of costs and cost risk before considering hosting the Games. The data presented in the present paper will allow such an assessment.

The IOC is not unaware of the issues. In 2015, Budapest, Hamburg, Los Angeles, Paris, and Rome submitted their first stage bids for hosting the 2024 Olympics. In 2015, Hamburg dropped out of the process after a referendum failed to garner support for the event (Livingstone 2015). In 2016, Rome withdrew because of the high level of debt it would create for the city (BBC 2016). In 2017, Budapest pulled out due to local opposition and a call to spend the money on health care and education instead (Wharton 2017). In the same year, the two cities remaining in the process, Paris and Los Angeles, were unconventionally awarded the 2024 and 2028 editions of the Games, without a proper bid process for the 2028 Games, for the simple reason that the IOC was running out of host cities. This development has been coming for a long time. Since Chicago's bid for the 2016 Olympics, organisers have faced increased scepticism about the claimed economic legacy and leverage of hosting the event (Lauermann 2019).

---

[1] Widely reported and used side-by-side, the original estimates do not include provisions for inflation, while the latest ones do. Below we remove those effects and compare like with like.

The unravelling of the bidding process culminated in 2017. It further fuelled the IOC's reform program, already underway. In 2015, the IOC had agreed on a program, the so-called Agenda 2020. The reforms continue today, now called Agenda 2020+5. The reforms aim to ensure the continued relevance of the Olympic Movement, safeguard the Olympic values, and strengthen the role of sports in society (Nicoliello 2021). In Agenda 2020, the only reform directly aimed at cost was a slight reduction of the effort required in the bidding process (Zimbalist et al. 2024), which we earlier dubbed "too little too late." (Flyvbjerg et al. 2021, p. 251). With Agenda 2020+5, the objective became "long-term sustainability, including from an economic standpoint." (IOC 2020, p. 7). The IOC now attempts to target cost savings by simplifying events, reusing solutions (e.g., ticketing), and avoiding the over-scoping of service levels and non-sports-related events. Whether those objectives are achievable remains to be seen and some observers already question them (Zimbalist et al. 2022). Below, we assess whether the reforms have so far reduced cost and cost overrun at the Games as was the intention, with Paris 2024 as a case in point.

# Cost of the Olympic Games 1960-2024

Table 1 shows the actual outturn costs of the Olympic Games 1960-2024 together with the number of events and number of athletes in each Games, in constant 2022 US dollars[2].

*Table 1 Outturn cost, number of athletes and events of the Olympic Games 1960-2024, constant 2022 US dollars.*

| Games Edition | Outturn Cost billions USD | Athletes | Events |
|---|---|---|---|
| *Summer Games* | | | |
| Rome 1960 | | 5,338 | 150 |
| Tokyo 1964 | 0.3 | 5,152 | 163 |
| Mexico City 1968 | | 5,516 | 172 |
| Munich 1972 | 1.1 | 7,234 | 195 |
| Montreal 1976 | 7.1 | 6,048 | 198 |
| Moscow 1980 | 7.7 | 5,179 | 203 |
| Los Angeles 1984 | 0.8 | 6,829 | 221 |
| Seoul 1988 | | 8,397 | 237 |
| Barcelona 1992 | 11.6 | 9,356 | 257 |
| Atlanta 1996 | 4.7 | 10,318 | 271 |
| Sydney 2000 | 5.2 | 10,651 | 300 |
| Athens 2004 | 3.1 | 10,625 | 301 |
| Beijing 2008 | 8.3 | 10,942 | 302 |
| London 2012 | 16.8 | 10,568 | 302 |
| Rio 2016 | 23.6 | 10,500 | 306 |
| Tokyo 2020* | 13.7 | 11,420 | 339 |
| Paris 2024** | 8.7 | 10,500 | 329 |
| *Winter Games* | | | |
| Squaw Valley 1960 | | 665 | 27 |
| Innsbruck 1964 | 0.0 | 1,091 | 34 |
| Grenoble 1968 | 1.0 | 1,158 | 35 |
| Sapporo 1972 | 0.1 | 1,006 | 35 |
| Innsbruck 1976 | 0.1 | 1,123 | 37 |
| Lake Placid 1980 | 0.5 | 1,072 | 38 |
| Sarajevo 1984 | | 1,272 | 39 |
| Calgary 1988 | 1.2 | 1,432 | 46 |
| Albertville 1992 | 2.1 | 1,801 | 57 |
| Lillehammer 1994 | 3.4 | 1,737 | 61 |
| Nagano 1998 | 2.2 | 2,176 | 68 |
| Salt Lake City 2002 | 2.7 | 2,399 | 78 |
| Torino 2006 | 4.7 | 2,508 | 84 |
| Vancouver 2010 | 3.2 | 2,566 | 86 |
| Sochi 2014 | 28.9 | 2,780 | 98 |
| Pyeong Chang 2018 | 3.4 | 2,833 | 102 |
| Beijing 2022 | 8.7 | 2,871 | 109 |

Note:
* Cost of USD 10.6 billion based on official accounts; see discussion below
** Paris costs are still in part estimates. The number of athletes is estimated from the cap stipulated in the host agreements

---

[2] Our approach to data collection and analyses is explained in Appendix B. The cost of Mexico City 1968 and Sarajevo 1984 are excluded due to hyperinflation, which results in implausible figures.

The mean cost for the Summer Games is USD 8.04 billion (median USD 7.38 billion, in 2022 prices). The Winter Games have a mean cost of USD 4.15 billion (median USD 2.18 billion, 2022 prices). The difference is statistically significant (Wilcoxon test, p = 0.015).

The most expensive Summer Games to date were Rio 2016 at USD 23.6 billion and London 2012 at USD 16.8 billion. The official Games Report puts the cost of Tokyo 2020/21 at USD 13.7 billion. Zimbalist et al. (2024) documented significant budget exclusions from official figures - "financial legerdemain," they call it (Zimbalist et al. 2024, p. 702). The Tokyo OCOG's costs exclude, for example, land costs and the City of Tokyo's budget. Adding those costs puts the cost of the Tokyo Games between USD 19.2 billion and 33.4 billion (2022 USD, excluding inflation). Zimbalist et al.'s 33.4 billion figure includes transport infrastructure and other investments, such as a weather satellite, which this study excludes. Even so, we can reasonably consider Tokyo the second most expensive, if not the most expensive, Summer Games in Olympic history.

For the Winter Games, Sochi 2014 was the most expensive at USD 28.9 billion, followed by Beijing 2022 at USD 8.7 billion. However, it is important to bear in mind that these figures do not include wider capital costs (OCOG indirect costs) for urban and transportation infrastructure, which are typically substantial. This omission raises questions about the true cost of hosting the Games and the potential implications of these rising costs, a topic that warrants further exploration.

Figure 1 presents a comprehensive view of the cost trends from 1964 to 2024. The trend lines clearly demonstrate a significant increase in the cost of hosting the Olympic Games over time. This finding is not only highly statistically significant overall (t(27) = 5.51, sandwich corrected p < 0.001), but also separately for Summer (t(12) = 3.68, sandwich corrected p = 0.005) and Winter Games (t(13) = 5.15, sandwich corrected p = 0.003). This upward trend, which was not statistically significant in our previous study (Flyvbjerg et al. 2016), is a new and important finding. It indicates that the Games have become increasingly expensive in today's money, a fact that should not be overlooked.

*Figure 1 Time series of outturn cost for Olympics 1964-2024 in billion 2022 USD, cost data shown as log10 and trends fitted for Summer and Winter Games using local polynomial regressions.*

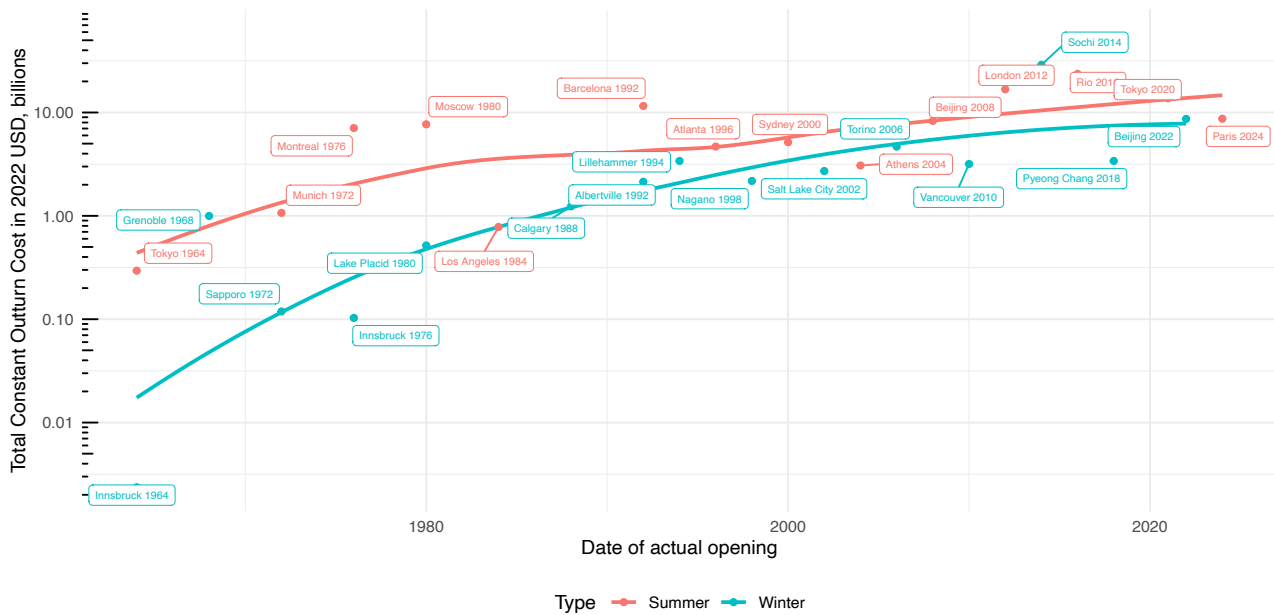

One of the changes in the IOC's Agenda 2020 is the shift from sports to events. The shift allows the IOC to engage in more granular discussions with the athletes and sporting federations about their requirements for each event that is part of the sport. For example, in Milano-Cortina 2026, the speed skating events will be held at the Fiera Milano Exhibition Centre. The choice of venue resulted from discussions between the IOC, OCOG, and the International Skating Unit – the governing body for competitive ice-skating disciplines. The IOC successfully pushed for a lower-cost solution than a purpose-built venue. Compare this to Beijing 2022. The specially constructed Ice Ribbon was the dedicated venue for the 166 athletes competing in the 14 speed skating events. The Ice Ribbon was estimated to cost USD 227 million (2022 USD), but the Beijing organisers have not disclosed the final actual cost of the venue (Teh and Stonington 2022). The high cost of the Olympic Games stems from the cost of each sporting discipline and event, which illustrates the difficulty of reducing costs, because each event must be negotiated.

Table 2 shows the cost per event and athlete from 1960-2024 in 2022 USD. The mean cost per athlete in the Summer Games 1960-2024 is USD 0.86 million (median USD 0.79 million), and in the Winter Games, USD 1.69 million (median USD 1.13 million). The cost per athlete is not statistically significantly different between the Summer and Winter Games (p = 0.683, Wilcoxon test).

The mean cost per event for the Summer Games is USD 28.7 million (median USD 27.0 million), while for the Winter Games, the mean is USD 49.1 million (median USD 33.3 million). Despite the difference, it's important to note that the costs are not statistically significantly different (p = 0.870, Wilcoxon test).

*Table 2 Cost per event and cost per athlete in the Olympics 1960-2024 in million 2022 USD.*

| Summer Games | Cost per Athlete (USD 2022, millions) | Cost per Event (USD 2022, millions) | Winter Games | Cost per Athlete (USD 2022, millions) | Cost per Event (USD 2022, millions) |
|---|---|---|---|---|---|
| Rome 1960 | NA | NA | Squaw Valley 1960 | NA | NA |
| Tokyo 1964 | 0.1 | 1.8 | Innsbruck 1964 | 0.0 | 0.1 |
| Mexico City 1968 | NA | NA | Grenoble 1968 | 0.9 | 28.5 |
| Munich 1972 | 0.1 | 5.5 | Sapporo 1972 | 0.1 | 3.4 |
| Montreal 1976 | 1.2 | 35.7 | Innsbruck 1976 | 0.1 | 2.8 |
| Moscow 1980 | 1.5 | 37.9 | Lake Placid 1980 | 0.5 | 13.6 |
| Los Angeles 1984 | 0.1 | 3.5 | Sarajevo 1984 | NA | NA |
| Seoul 1988 | NA | NA | Calgary 1988 | 0.9 | 26.8 |
| Barcelona 1992 | 1.2 | 45.0 | Albertville 1992 | 1.2 | 37.5 |
| Atlanta 1996 | 0.5 | 17.3 | Lillehammer 1994 | 2.0 | 55.7 |
| Sydney 2000 | 0.5 | 17.2 | Nagano 1998 | 1.0 | 32.0 |
| Athens 2004 | 0.3 | 10.2 | Salt Lake City 2002 | 1.1 | 34.8 |
| Beijing 2008 | 0.8 | 27.5 | Torino 2006 | 1.9 | 55.7 |
| London 2012 | 1.6 | 55.6 | Vancouver 2010 | 1.2 | 36.9 |
| Rio 2016 | 2.3 | 77.2 | Sochi 2014 | 10.4 | 295.1 |
| Tokyo 2020 | 1.2 | 40.5 | Pyeong Chang 2018 | 1.2 | 33.3 |
| Paris 2024 | 0.8 | 26.5 | Beijing 2022 | 3.0 | 79.6 |

Figures 2 and 3 show the cost per athlete over time. Statistical tests show that Sochi 2014 and Beijing 2022 are extreme values in their cost per athlete. Sochi 2014 cost USD 10.4 million per athlete. Beijing 2022 had a cost of USD 3.0 million (both in 2022 terms).

Figure 3 shows the same data with the two extreme values removed. This is to test whether the trend towards higher cost is robust, which it is. Without the extremes, there is still a statistically significant trend over time for higher costs, both overall (t(25) = 3.95, sandwich corrected p = 0.007), for the Summer Games separately (t(12) = 2.56, sandwich corrected p = 0.040), and for the Winter Games separately (t(11) = 3.19, p = 0.009). This upward trend, which again was not statistically significant in our previous study (Flyvbjerg et al. 2016), is a new and important finding.

The analyses show that the Games have become increasingly expensive over time measured in today's money, both overall and for each type of Games, and both measured per athlete and per event.

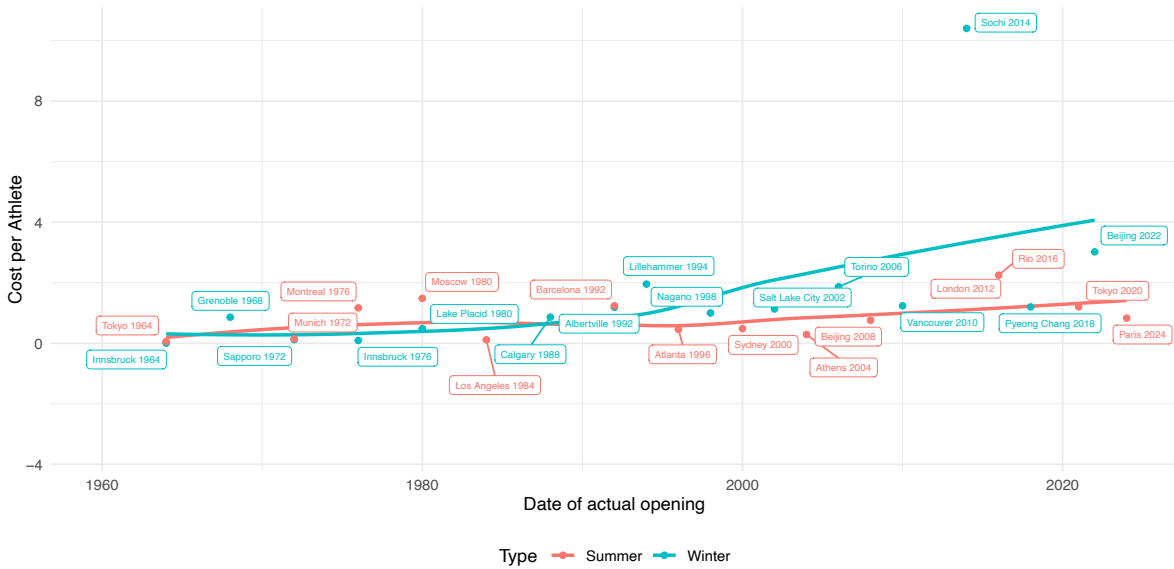

*Figure 2 Cost per athlete in million 2022 USD for the different editions of the Olympic Games, cost data shown as log10 and trends fitted for Summer and Winter Games using local polynomial regressions.*

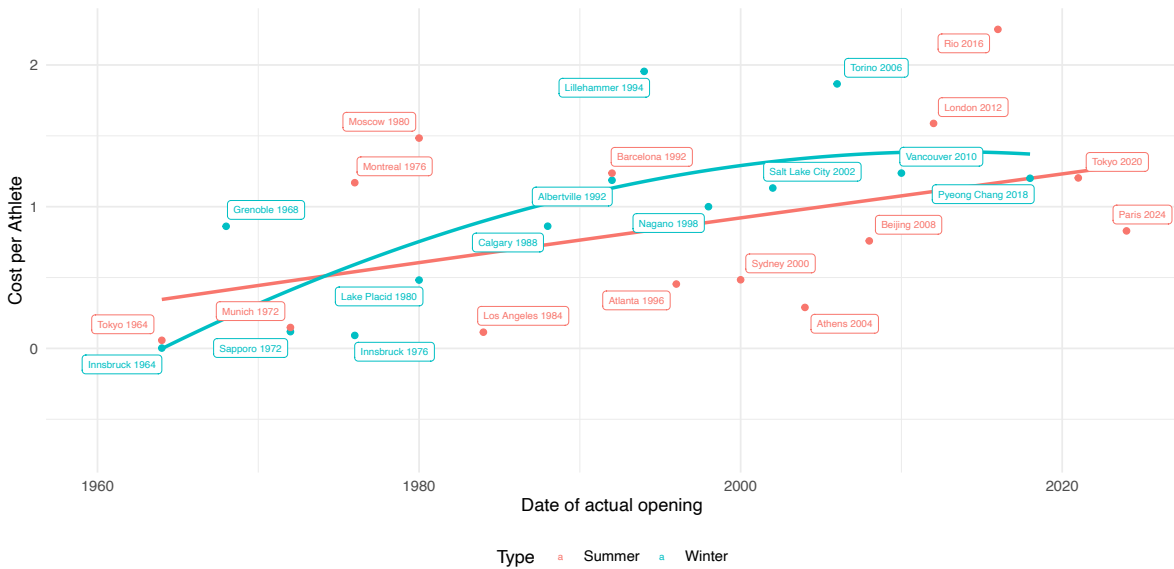

*Figure 3 Cost per athlete in million 2022 USD of the different editions of the Olympic Games, cost data shown as log10 and trends fitted for Summer and Winter Games using local polynomial regressions, extreme values removed.*

# Cost Overrun at the Games 1960-2024

Tables 3 and 4 show cost overrun for the Games, which is a key interest for our research. Overruns are measured in real terms, i.e., with inflation removed, and in local currencies (see Appendix B). Table 3 presents an overview while Table 4 provides the detailed figures for each Games. For cost overrun in real terms, the numbers document a systematic and highly significant tendency to overrun (overruns are statistically significantly greater than 0%, Wilcoxon test, $p < 0.001$). Judging from these statistics it is clear that large risks of large cost overruns are inherent to the Olympic Games.

Despite the Winter Games being smaller in scope than the Summer Games, their cost risks are not statistically significantly different (Wilcoxon, real-term overruns, $p = 0.605$). This suggests that the scale or complexity of the Games is unlikely to be the sole explanation of overrun. This finding challenges conventional wisdom and opens up new avenues for research and discussion.

For cost overrun in nominal terms, these are the figures that typically make the headlines in media reports. We stress, however, that nominal cost, unlike real cost, does not present a like-for-like comparison. The candidature files do not include a provision for anticipated inflation. Estimates are in real terms, usually at a price level a year before the host selection. Subsequent budget announcements often - but not always - include inflation, and reported outturn costs all include inflation. In other capital projects, providing for inflation in forecasting is standard practice. The planners for the Games have yet to do the same. These numbers are thus only correct if we take the candidature files at face value, which means that the hosts did not anticipate any inflation between the estimate and the expenditure. Again, this is how bidders misrepresent their estimates, making them look low.

*Table 3 Mean and median cost overruns of the Olympic Games 1960-2024.*

|  | Real terms | | Nominal terms | |
| --- | --- | --- | --- | --- |
|  | **Mean** | **Median** | **Mean** | **Median** |
| **Summer** | 195% | 121% | 336% | 142% |
| **Winter** | 132% | 118% | 277% | 169% |
| **Overall** | 159% | 118% | 302% | 146% |

From Table 4, which shows cost overrun in nominal and real terms for all Games with available data, we further observe:

- All Games (100%) have cost overruns,
- 18 of 23 Games (78 percent) have cost overruns above 50 percent in real terms, and
- 13 of 23 Games (57 percent) have cost overruns above 100 percent in real terms.

*Table 4 Cost overruns of the Olympic Games 1960-2024.*

| Summer Games | Nominal | Real | Winter Games | Nominal | Real |
| --- | --- | --- | --- | --- | --- |
| Rome 1960 | NA | NA | Squaw Valley 1960 | NA | NA |
| Tokyo 1964 | NA | NA | Innsbruck 1964 | NA | NA |
| Mexico City 1968 | NA | NA | Grenoble 1968 | 230% | 181% |
| Munich 1972 | NA | NA | Sapporo 1972 | NA | NA |
| Montreal 1976 | 1266% | 720% | Innsbruck 1976 | NA | NA |
| Moscow 1980 | NA | NA | Lake Placid 1980 | 502% | 324% |
| Los Angeles 1984 | NA | NA | Sarajevo 1984 | 1257% | 118% |
| Seoul 1988 | NA | NA | Calgary 1988 | 131% | 65% |
| Barcelona 1992 | 609% | 266% | Albertville 1992 | 169% | 137% |
| Atlanta 1996 | 178% | 151% | Lillehammer 1994 | 347% | 277% |
| Sydney 2000 | 108% | 90% | Nagano 1998 | 58% | 56% |
| Athens 2004 | 97% | 49% | Salt Lake City 2002 | 40% | 24% |
| Beijing 2008 | 35% | 2% | Torino 2006 | 113% | 80% |
| London 2012 | 108% | 76% | Vancouver 2010 | 36% | 13% |
| Rio 2016 | 673% | 352% | Sochi 2014 | 508% | 289% |
| Tokyo 2020 | 139% | 128% | Pyeong Chang 2018 | 14% | 2% |
| Paris 2024 | 146% | 115% | Beijing 2022 | 190% | 149% |

For the Summer Games the largest cost overrun was found for Montreal 1976 at 720 percent in real terms, followed by Rio 2016 at 352 percent. The smallest cost overrun for the Summer Games was found for Beijing 2008 at two percent, followed by Athens 2004 at 49 percent. The last three Summer Games all had cost overruns greater than 100%, Rio 2016, Tokyo 2020, and Paris 2024 more than doubled their budgets.

For the Winter Games, the largest cost overruns are Lake Placid 1980 at 324 percent in real terms followed by Sochi 2014 at 289 percent. The smallest cost overrun for the Winter Games was found for Pyeong Chang 2018 at 2 percent, followed by Vancouver 2010 at 13 percent.

Again, we emphasise these are conservative figures. Observers of the Games have indicated that the OCOGs' final accounts are likely to understate the true cost of the Games, e.g., Tokyo 2020 (Zimbalist 2020, Zimbalist et al. 2024) and Beijing 2008 (Flyvbjerg et al. 2016, Flyvbjerg et al. 2021). Thus, the true overruns are likely to be higher.

Table 5 compares the overruns of the Olympics with other capital projects. The data further stress the financial and economic risk of hosting the Olympic Games because:

1. *All Games, without exception, have cost overrun*. For no other type of megaproject is this the case, not even the construction of nuclear power plants or the storage of nuclear waste. For other capital investment types, typically 10-60 percent of investments come in on or under budget. For the Olympics, it is zero percent. It is worth considering this point carefully. A budget is typically established at a reasonable maximum value to be spent on a capital investment. However, in the Games the budget is more like a fictitious minimum that was never sufficient. Further, the host guarantees that they will cover the cost overruns of the Games. Flyvbjerg et al. (2016) suggest that this guarantee is akin to writing a blank check for the event, with certainty that the cost will be more than what has been quoted. We called this the "Blank Check Syndrome." In practice, the bid budget is on average a 38% down payment; further instalments will follow, written on the blank check.
2. *The Olympics have the second highest average cost overrun of any type of megaproject*, at 159 percent in real terms. Only nuclear waste disposal projects have higher overrun at 238% in real terms.

To compare, nuclear power plants have the highest mean cost overrun after the Olympics at 120 percent. Despite frequent objections, the Games have nowhere near the degree of local opposition that is typical for nuclear waste storage and nuclear power plants. Rather, the high cost overruns for the Games may be related to the fixed deadline for delivery and fixed scope. The opening date cannot be moved and scope is negotiated directly between the Athletes' Federations and the IOC, neither of whom have skin in the Game for any overspend. Therefore, when problems arise there can be no trade-off between schedule, scope and cost, as is common for other megaprojects. All that OCOGs can do is to allocate more money, which is what happens. This is the Blank Check Syndrome, again.

3. *The high average cost overrun for the Games, combined with the existence of extreme values, should be cause for caution for anyone considering hosting the Games*, and especially small or fragile economies with little capacity to absorb escalating costs and related debt (Flyvbjerg et al. 2021). A one-in-five risk of a 50+ percent cost overrun of the Olympics should concern government officials and taxpayers. Such overrun may have fiscal implications for decades to come, as happened with Montreal where it took 30 years to pay off the debt incurred by the 720 percent cost overrun on the 1976 Summer Games (Vigor et al., 2004: 18), and Athens 2004 where Olympic cost overruns and related debt exacerbated the 2007-17 financial and economic crises, as mentioned above (Flyvbjerg, 2011).

*Table 5 Comparison of the cost overruns at the Olympics and other types of capital projects (real terms, local currencies).*

| Project Type | n | Mean cost overrun | Percentage of projects on budget or below | Percentage of projects with cost overrun > 50% |
|---|---|---|---|---|
| Nuclear Storage | 23 | 238% | 9% | 48% |
| Olympics | 23 | 159% | 0% | 78% |
| Nuclear Power | 196 | 120% | 3% | 55% |
| IT | 5403 | 73% | 59% | 18% |
| Dam | 334 | 71% | 26% | 35% |
| Aerospace | 97 | 60% | 9% | 42% |
| Building | 310 | 52% | 29% | 25% |
| Defence | 132 | 49% | 49% | 18% |
| Nuclear Decommissioning | 29 | 48% | 21% | 38% |
| Rail Station | 70 | 43% | 30% | 24% |
| Rail | 540 | 34% | 29% | 25% |
| Oil and Gas | 101 | 32% | 19% | 18% |
| Stadium | 51 | 32% | 25% | 16% |
| Mining | 886 | 27% | 51% | 16% |
| Healthcare Infrastructure | 102 | 19% | 50% | 11% |
| Steel Mill | 19 | 19% | 11% | 5% |
| Thermal Power | 189 | 18% | 44% | 13% |
| Road | 2250 | 16% | 42% | 9% |
| Water | 345 | 15% | 40% | 11% |
| Pipeline | 412 | 14% | 45% | 6% |
| Wind Power | 82 | 13% | 45% | 7% |
| Pumped Hydro | 36 | 11% | 39% | 8% |
| Battery Storage | 10 | 6% | 40% | 0% |
| Energy Transmission | 54 | 5% | 63% | 4% |
| Solar Power | 41 | 1% | 61% | 0% |

Source: Oxford Projects Database Q2 2023

# Have Cost Overruns Decreased?

In Flybjerg et al. (2016) we examined the impact of the Olympic Knowledge Sharing Program, which was first used in Sydney 2000.[3] We observed a decrease in cost overruns up to 2008 and saw anecdotal evidence of increasing overruns after 2008. Now, with more data, we can statistically test this pattern.

Figure 4 shows historical trends with cost overrun on a log scale. We pooled the data because cost overrun is not statistically significantly different between the Summer and Winter Games, as we saw above. The data indicate a convex trend: Until 2008 cost overruns decreased and from 2010 it increased. Figure 5 formally tests this impression with a segmented linear fit; the algorithm identifies 2008 as the most likely change point with the change trends being statistically significant (bootstrapped p-values all < 0.02).

The data show, with statistical significance, that the Knowledge Sharing Program failed to reduce cost and cost overrun at the Games. More fundamental interventions are needed.

*Figure 4 Cost overruns for the Olympic Games 1968-2024 in real terms, local currencies, cost data shown as log10 and trend fitted using local polynomial regressions.*

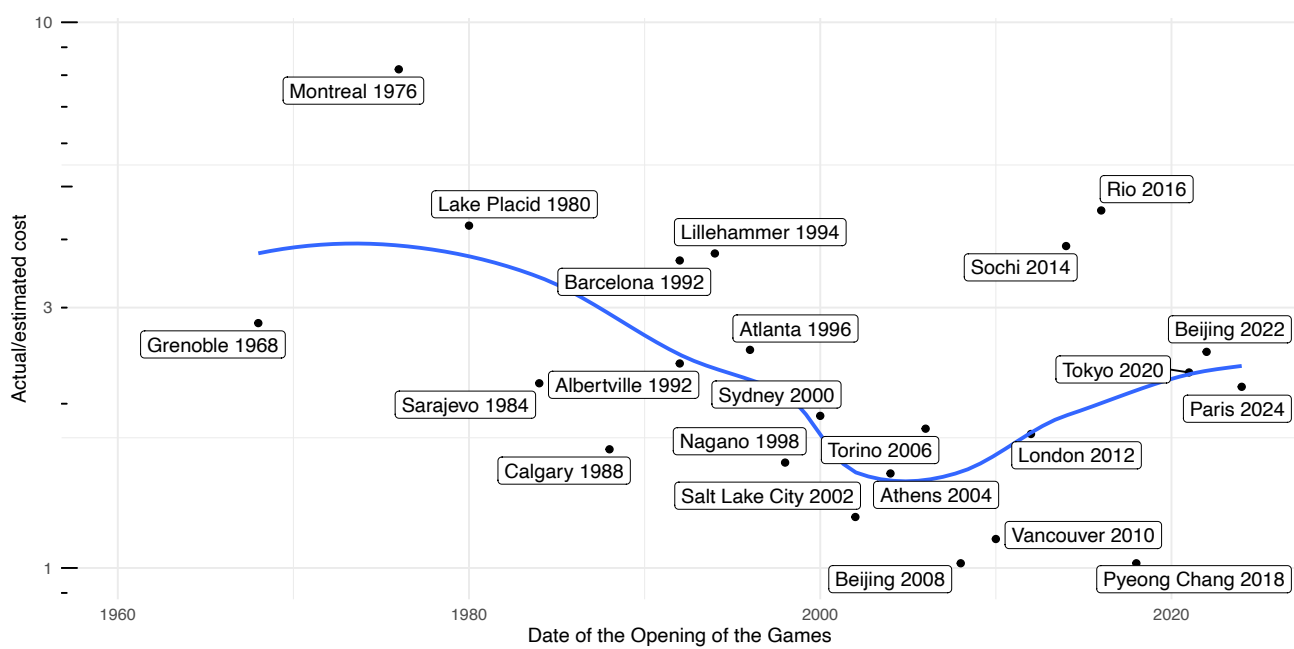

---

[3] The Olympic Knowledge Sharing Program began informally with Sydney 2000. It was launched officially in 2003 as the Olympic Games Knowledge Management (OGKM) program. In 2020, the OGKM program was merged with the Information and Knowledge Management (IKM) unit at the IOC, and became a team called Information, Knowledge, and Games Learning (IKL). The team refers to their activities as the "Knowledge Sharing Program" (source: private conversation with IOC officials, May 2023).

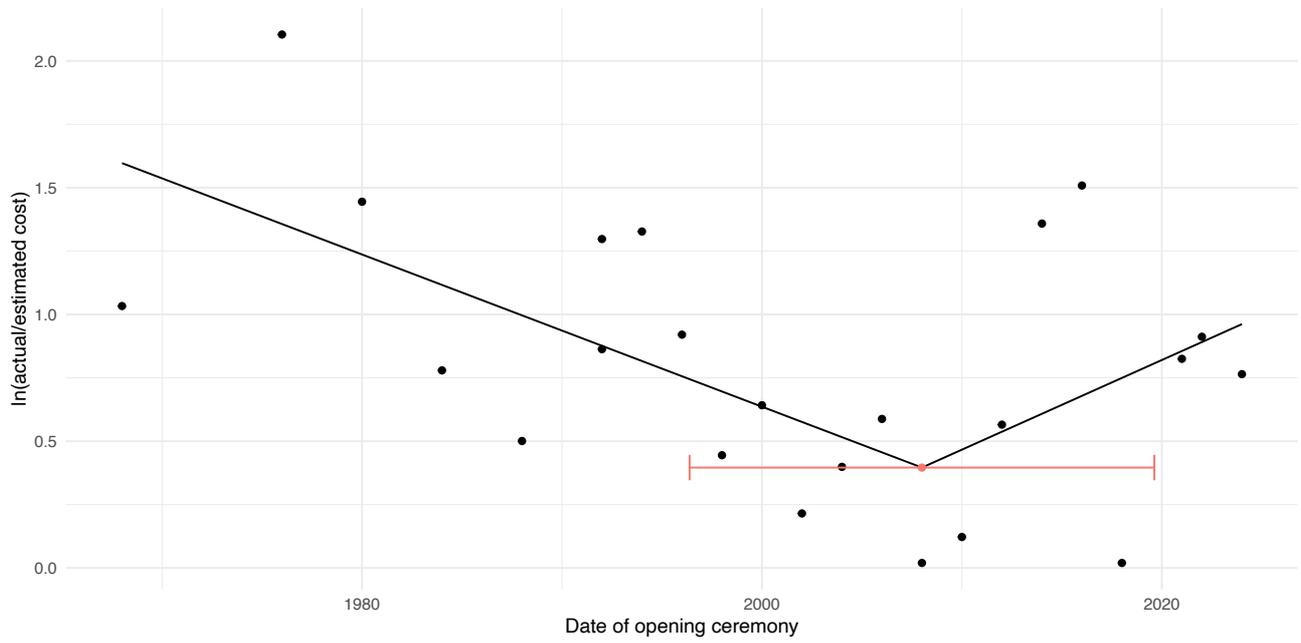

*Figure 5 Segmented fit and change-point analysis of the cost overruns of the Olympic Games 1968-2024 (estimated change point and 95% confidence interval of the change point in red using Delta method).*

# Paris 2024

Paris was selected for the 2024 Games after a conventional bid that went awry because of several potential hosts withdrawing, as mentioned above. Importantly, Paris 2024 is the first Games to extensively follow the reuse and retrofit policy of Agenda 2020+5. The only new construction for Paris 2024 is the Aquatics Centre and the Athlete's Village (IOC 2024). We wanted to test whether the new policy has impacted cost and cost overrun.

The latest cost estimate for Paris 2024 is USD 8.7 billion (2022 prices, real terms). This is USD 1.32 billion more than the historic median for the Summer Games. It is lower, however, than the three most recent Summer Games (Tokyo 2020/21, Rio 2016, London 2012), while it is higher than the three Games before that (Beijing 2008, Athens 2004, Sydney 2000), as Figure 6 shows.

It may be that the shift to reuse/retrofit facilities has contributed to the lower costs for Paris 2024 compared with the three previous Summer Games. But given that Paris 2024 is not in the low end of cost for all Games, and given that the Paris costs are still an estimate that could end up higher, this conclusion is not particularly convincing.

Moreover, so far Paris 2024 has had a 115% cost overrun in real terms, which places them in the middle range of previous Summer Games, with a substantial risk of further overrun (see Figure 7).

In conclusion, costs have decreased for Paris 2024 compared with the three most recent previous Summer Games. But Paris costs are still just an estimate, they are above other recent Games, and they are more than a billion USD above the historical median cost. Moreover, the Paris decrease in cost has not happened at the rate expected, which is attested by the cost overrun incurred. Paris 2024 is a first, however, in terms of reuse/retrofit and perhaps it will take time before the new policy matures and becomes effective, meaning that future Games could have lower costs than Paris. The IOC will certainly hope so in order to give potential host cities less reason to walk away from the Games due to high costs.

*Figure 6*

*Comparison of estimated cost for Paris 2024 (red, estimate) with actual outturn cost of previous Games 1990-2022 (black), billion 2022 USD, real terms.*

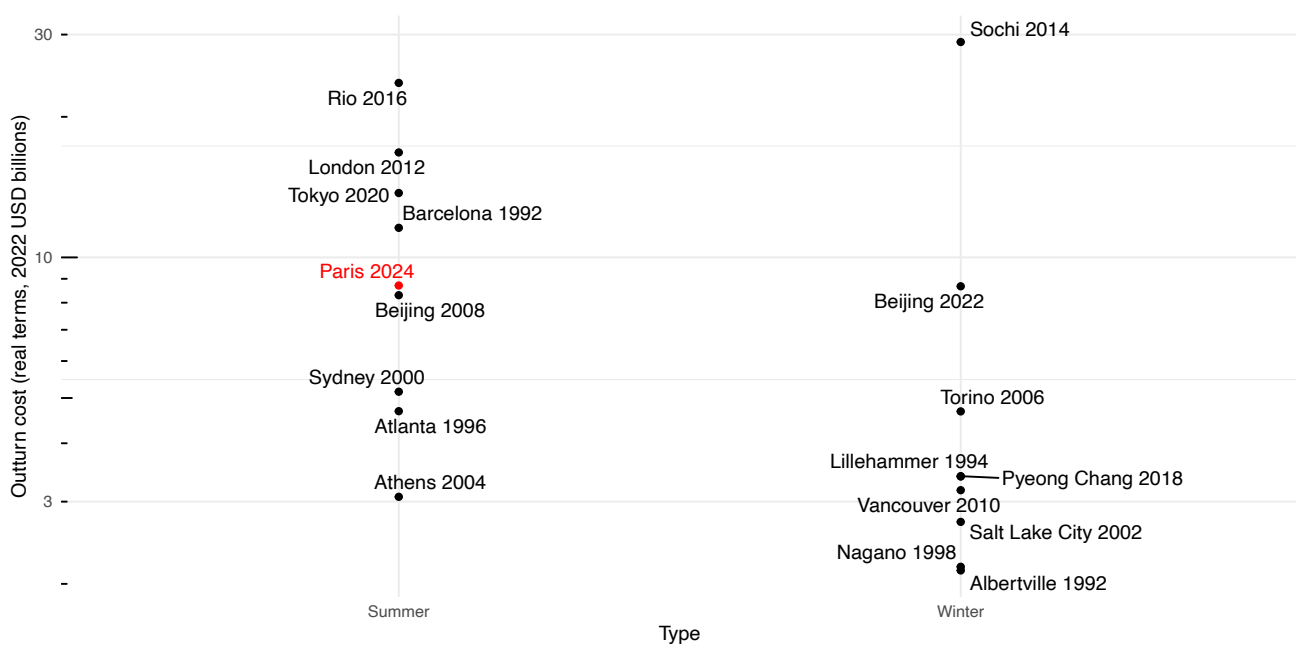

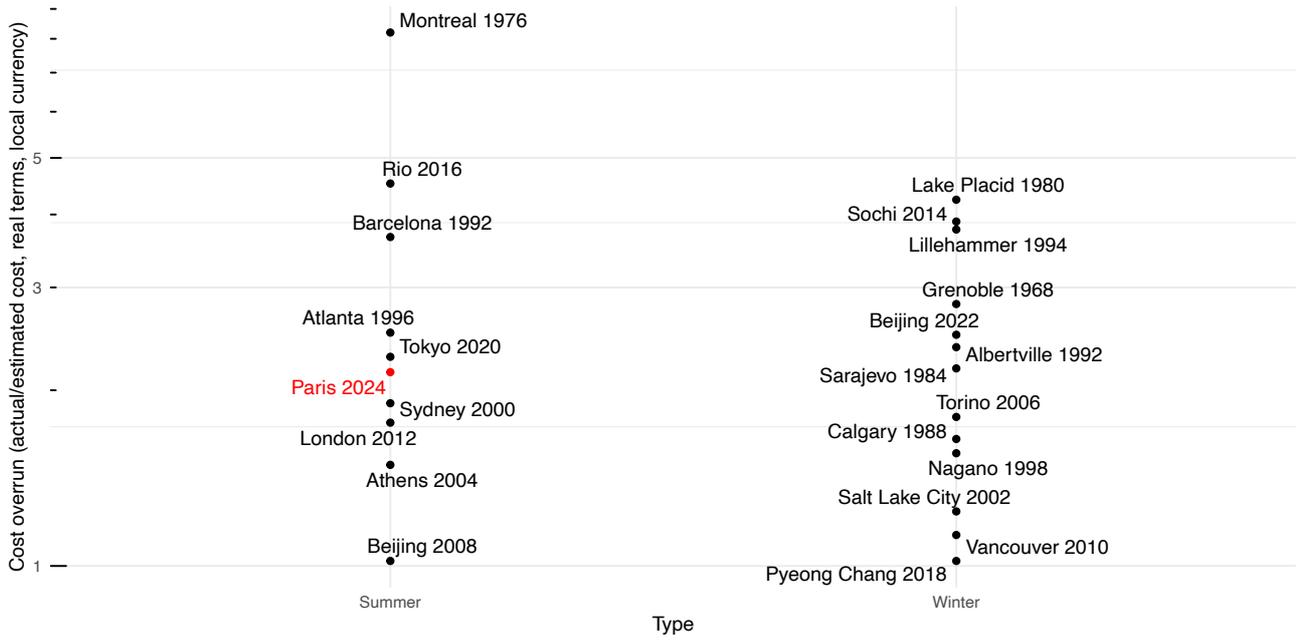

Figure 7 Comparison of cost overrun for Paris 2024 to date (red, estimate) with actual cost overrun for previous Games (black), real terms.

# Future Games 2026-2034

The next Winter Games are Milano-Cortina 2026, the French Alps 2030, and Salt Lake City-Utah 2032. The next two Summer Games are Los Angeles 2028 and Brisbane 2032. Table 6 shows the latest cost estimates for these four Games.

Milano-Cortina 2026 shows the same pattern as Paris 2024 in the sense that (a) the estimated cost (2.6 billion 2022 USD) is higher than the historic median for all Winter Games by USD 420 million and (b) it is lower than the cost of the most recent Winter Games in Beijing 2022 (USD 8.7 billion), Pyeong Chang 2018 (USD 3.4 billion), and Sochi 2016 (USD 28.9 billion), but (c) higher than Nagano 1998 (USD 2.2 billion) and Albertville 1992 (USD 2.1 billion). Similarly, current estimates for French Alps 2030 and Salt Lake City-Utah 2034 are higher than the historic median but lower than recent extremes.

The similarity of Milano-Cortina with Paris continues with a large cost overrun, of 78% in real terms, so far. This places Milano-Cortina in the middle range of previous Winter Games regarding cost overrun. Again, it must be remembered that the cost for Milano-Cortina 2026 is an estimate that could end up significantly higher, that is, with a higher cost overrun than the 78% recorded to date.

*Table 6 Latest cost estimates and cost overruns of the Games 2026-2032. Estimates and overruns are likely to increase in the future.*

| Games | Latest estimate in billion 2022 USD | Cost overrun nominal terms | Cost overrun real terms |
|---|---|---|---|
| Milano-Cortina 2026 | 2.6 | 102% | 78% |
| Los Angeles 2028 | 5.9 | 28% | -5% |
| French Alps 2030 | 2.8-3.5 | - | - |
| Brisbane 2032 | 3.6 | 44% | 2% |
| Salt Lake City-Utah 2034 | 3.5 | - | - |

Further into the future we find Los Angeles 2028 and Brisbane 2032. Both Games have had escalating budgets, Los Angeles from USD 5.3 billion to USD 6.8 billion, Brisbane from AUD 4.9 billion (USD 3.8 billion) to AUD 7.1 billion (USD 4.7 billion). Both of the latest budgets include inflation provisions, while the original budgets did not.

In Table 6, we used IMF and OECD forward inflation projections to calculate cost overrun incurred so far in real terms. Removing inflation shows that Los Angeles could be the first Games ever to deliver under budget (5% under budget in real terms). But this would only happen if the estimated budget were to remain unchanged and US construction inflation to not run higher than the 2.6% forecasted by the OECD. Both of those assumptions are big ifs when we look at the history of persistent cost overrun for the Games and persistent construction cost inflation above other inflation. Similar considerations apply even more to Brisbane 2032, where, unlike Los Angeles, organisers are still debating venue options and their possible costs (Messenger 2024). Such fundamental uncertainties make it likely that Brisbane will see further increases in cost and cost overrun.

While the outturn costs of the future Games are highly uncertain, we do know the estimated bid cost for the full period 1960 to 2032, more than 70 years total. Figure 8 shows those numbers. We emphasize that these are all estimated cost and note an increase in bid budgets up to 2010 followed by a decrease. The regression analysis illustrated in Figure 9 finds the change in trend to be statistically significant (all $p < 0.001$).

But what could possibly explain this marked shift in trend?

First, there might be a political bias in the data. Agenda 2020 intended to decrease the cost of the Games, but without any real changes, which were only introduced with Agenda 2020+5. Zimbalist (2020) and Zimbalist et al. (2024) document how the IOC put pressure on the Tokyo 2020/21 organizers to reduce the publicly stated budgets below expert cost forecasts. Cost overrun followed cost underestimate, as always.

Second, optimism bias may also be present in the data. The budgets of both Los Angeles 2028 and Brisbane 2032 are based on optimistic assumptions of (a) low future inflation and (b) no further scope changes. History does not support these assumptions. Similarly, optimism is likely also present in assumptions of how effective the reuse/retrofit policy will be. The estimates assume their effectiveness. But as we saw for Paris 2024 that assumption may not hold up.

If the decrease in estimated cost after 2010 is indeed caused by such biases then the future is likely to bring more cost overruns at the Games, not less, because overruns follow underestimates as surely as day follows night.

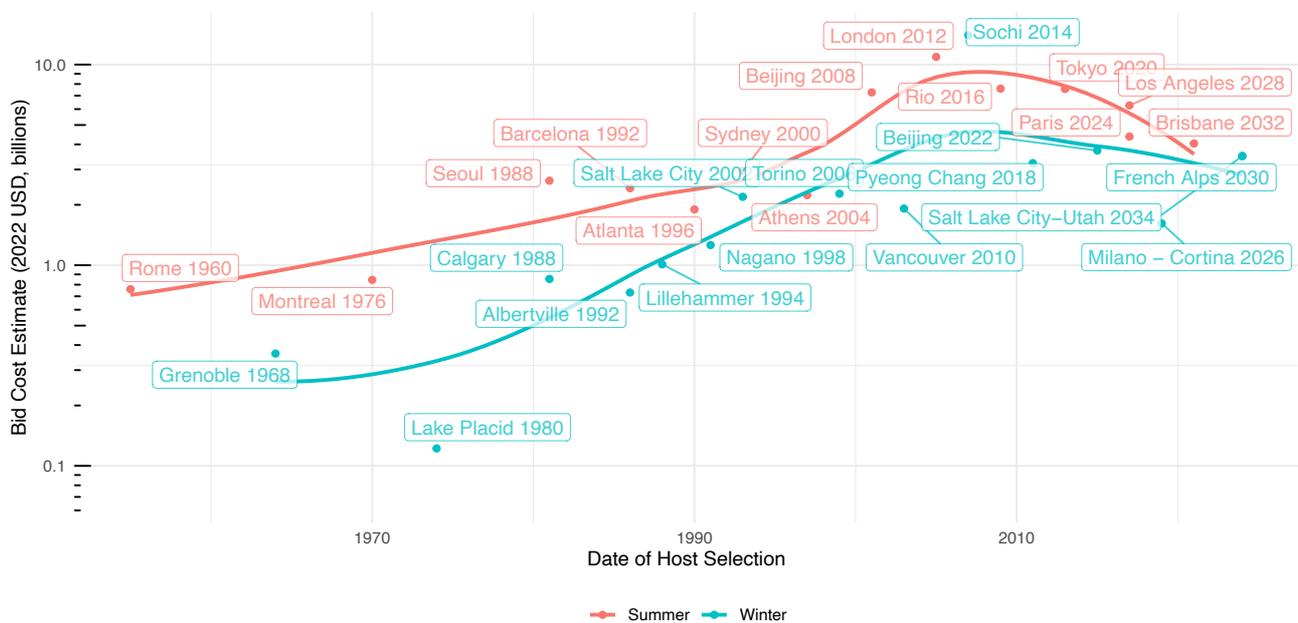

*Figure 8 Bid cost estimate of the Olympics 1960-2032 in billion 2022 USD, cost estimate data shown as log10 and trends fitted for Summer and Winter Games using local polynomial regressions.*

*Figure 9 Change-point analysis and segmented fit of the bid cost estimates for the Summer and Winter Games 1960-2032 (estimated change points and 95% confidence interval of the change points in red using delta method.*

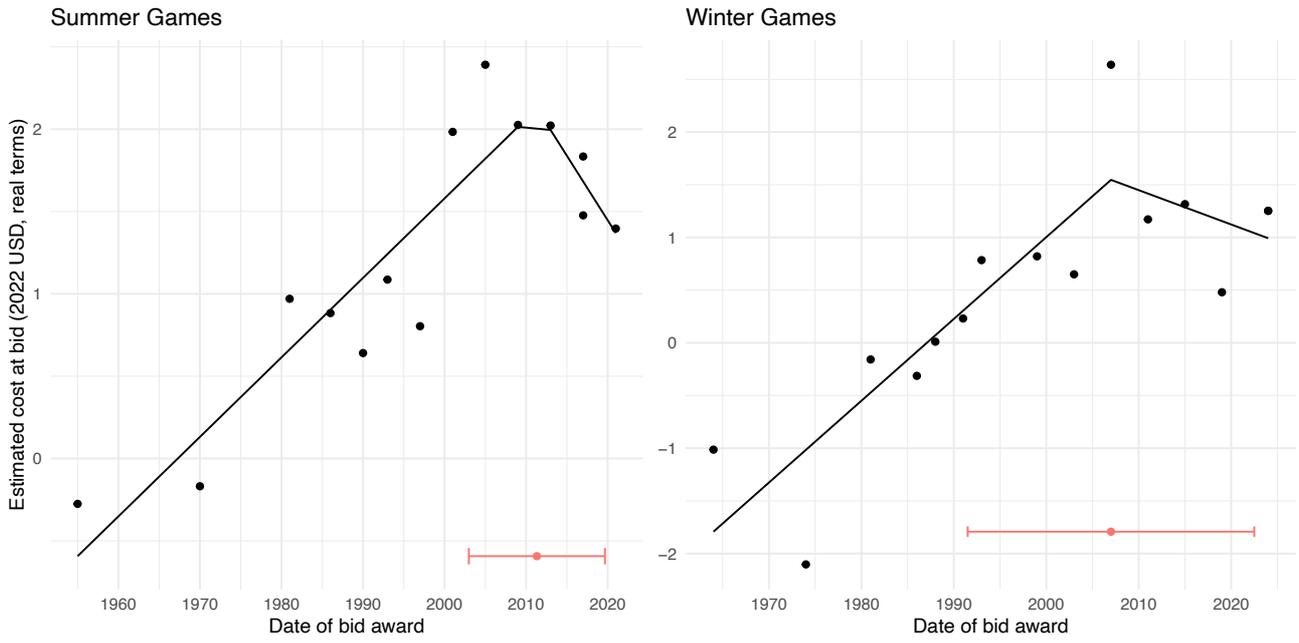

# What More Can Be Done?

After many years of debate, finally, there seems to be a growing consensus that the Olympics tend to be expensive and run over budget. There is also increasing agreement regarding what the numbers for cost and cost overrun are for each Games. Now that these facts are no longer swept under the carpet, we can begin to focus on improving the situation.

A first improvement would be better planning. Our longstanding advocacy for the use of historical data in decision-making is rooted in its ability to enhance the quality of decisions by reducing bias (Flyvbjerg 2006). Kahneman et al. (2021, pp. 326-27) call this 'decision hygiene' and it comes highly recommended with solid proof of its efficacy. However, this practice is contingent upon the availability of valid data. Historically, unfortunately neither host cities nor the IOC have been conducive to producing such data. Sometimes quite the opposite. For example, the Rio 2016 Games have yet to submit their final accounts, while Tokyo 2020/21 has provided data that align better with the true cost than pre-event estimates, though uncertainties persist for Tokyo, too.

Enhancing the quality of data necessitates a commitment to transparency, which has not traditionally been part of IOC culture. While some progress has been made in this area, the inclination to present underestimated costs persists, as we saw above. To truly improve, complete costs must be shared, encompassing private contributions and non-OCOG budgets. The IOC should take the lead in this by transparently capturing data on the cost of the Games. This is particularly important given the potential cost savings from venue reuse/retrofit and the use of temporary venues, which the IOC recommends but which have yet to be quantified and monitored.

Second, the new practice of non-committed dialogue for selecting host cities has extended the preparation time for the Games. Los Angeles and Brisbane, for instance, have 12 years to prepare compared to the usual 8 years. Our study underscores the significant challenge that inflation poses for projects of this magnitude over such a long time horizon. Given that provisions for inflation are standard practice for other megaprojects, it should also be a mandated component of future Games budgets, or the budgets will continue to be misleading, and the more misleading the longer the planning period, other things being equal.

Third, planners need to be more realistic about the uncertainties that projects as large as the Olympics face. Figure 6 shows the cost and cost overrun distributions for the Olympic Games 1968-2024. Half of the Games had a cost overrun of more than 118% in real terms, the other half less than that. For planning purposes, this means that if the funder of the Games had a risk appetite that accepts a 50% chance of overrun, then a 118% contingency would need to be added to the bid cost estimate to arrive at a realistic budget, still with a 50% risk of going over that budget. If the funder had a lower risk appetite, say 20%, then 273% would need to be added, as a historical fact.

This approach to using historical data assumes that the next Olympics will perform like previous ones. While this is not the intention of anyone involved, history shows that unfortunately it is a prudent assumption for forecasters and decision makers. Other megaprojects use something called the "20-50-70 approach": They evaluate the economic benefits case for their project at a 20% risk appetite (resulting in a 273% contingency for the Games), they budget for the project at a 50% risk appetite (118% contingency), and finally set a stretch target for the delivery team, to put pressure on them to perform, at a 70% risk appetite (72% contingency). The Olympics need to learn from approaches like this, which are standard elsewhere for project types that have been forced to take cost and cost overrun more seriously than the Games.

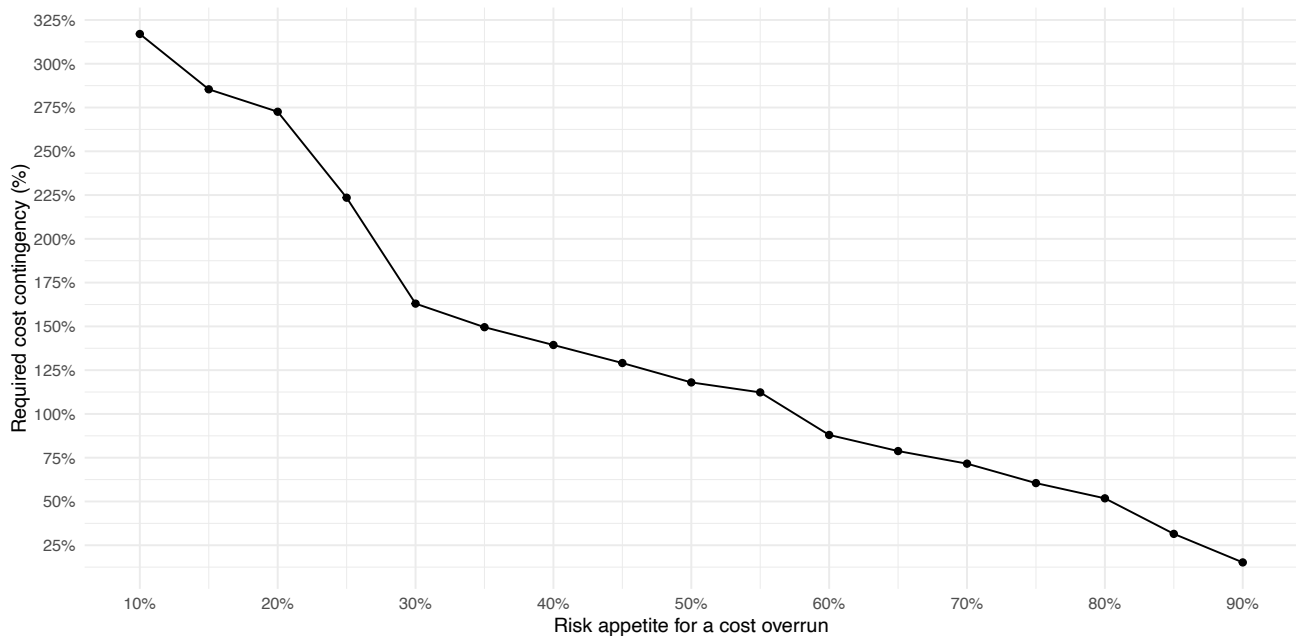

*Figure 10 Reference class forecast of the cost risk of the Olympic Games, real terms.*

These realistic contingencies exceed the typical 10-15% in the Games' budgets, which history has proven to be are fairy-tale optimistic, over and over. The first big question that potential hosts need to ask of the Games is therefore: Do the benefits make economic sense even if costs quadruple? If not, how can we increase the benefits?

A further obvious question is: How can we overcome history and outperform historical costs and cost overruns? Our analysis shows that reuse/retrofit of existing venues might reduce costs, but so far not to the degree anticipated by budgets. To bring realism into budgets, we need detailed sharing of data and experience about the actual cost savings from reuse and retrofit. Realistic budgets will also help reduce cost overruns. Emerging data show that the cost savings fall short of targets to make the Games affordable. Additional ideas and more work are needed.

Agenda 2020+5 established a focus on events. These may be considered the Games' modular building blocks (Flyvbjerg 2021, Flyvbjerg and Gardner 2023). Positive learning curves from reusing and repeating modules, over and over, are needed for effective improvement, both at the level of the building blocks (the events) and at the level of integrating all necessary building blocks (the Games).

The biggest objection we hear to reusing and repeating modules is the mental image of prefabricated 1970s-1980s housing projects. But nothing could be further from the truth with today's technology, materials, and build quality. An exciting challenge is how to make each edition of the Games iconic at significantly lower cost and cost overrun than we have seen to date. This can be done, but for it to happen the IOC and hosts need to change their tack and become more innovative in their approach to delivering the Games.

# Acknowledgements

The authors wish to thank Allison Stewart, University of Oxford, for commenting on an earlier draft of the paper; Giuseppe Sassano, UCL, for supporting data collection; and Mariagrazia Zottoli, University of Oxford, for checking the statistical analyses.

# Author Statement

The authors are listed alphabetically based on equal contributions to the paper.

# Appendix A: Previous Studies

Flyvbjerg and Stewart (2012) conducted the first systematic study of cost and cost overrun at the Olympic Games, expanding on previous work by Chappelet (2002), Essex and Chalkley (2004), and Preuss (2004). This research was further extended in Flyvbjerg et al. (2016) and Flyvbjerg et al. (2021). Today, cost studies of the Games attract sustained and broad attention in both research, policy, and practice.

Since 2012, additional studies problematized Olympic costs and cost overruns (e.g., Zimbalist 2015, Baade and Matheson 2016). Research has also documented the cost and overruns of Rio 2016 (Zimbalist 2016) and Tokyo 2020/21 (Waldenberger 2020). These studies show a continuation of the "high cost and large cost overrun" pattern documented in Flyvbjerg et al. (2021), Flyvbjerg et al. (2016), and Flyvbjerg and Stewart (2012).

In Flyvbjerg et al. (2021), we discovered that a Pareto distribution with infinite mean and variance best describes cost overrun for the Games. Mathematically-statistically, the power law explains why the Games are so difficult to plan and manage successfully: with infinite mean and variance conventional forecasting quite simply does not work. Substantively, we explained this convex outcome by the irreversibility of the hosting decision; the fixed deadlines that apply to the Games unlike to other project types;[4] something we called the "Blank Check Syndrome;" tight coupling; long planning horizons; and finally, an "Eternal Beginner Syndrome."

Preuss et al. (2019 and previously Preuss 2004) contained the first economic analyses of multiple Games. Preuss found that since 1972, every Organizing Committee of the Olympic Games (OCOG), which leads the planning of the Games in the host city, has produced a balanced budget. However, the study excluded sports-related investments such as venues. Others, including us, have argued that such investments should be included in cost analyses of the Games (Müller et al. 2021).

Recently, the debate has converged. Preuss and Weitzmann (2023, p. 471) now conclude, using more extensive sources than OCOG budgets, that "cost overruns occur at all Games". However, the reasons for overruns and their sizes are still contested (Weitzmann and Preuss 2023).

The debate over cost and cost overrun has further led to questions of sustainability. Cost is an essential indicator of the economic sustainability of the Games. Müller et al. created a database of megaevents covering the Olympics and Football World Cups (2022a), which found that these megaevents suffer from a persistent and systematic negative return on investment independent of the local context of decisions, project delivery, and economic cycles (Müller et al. 2022b). Müller et al. also questioned the Games' ecological, social, and economic sustainability. Their data show a declining trend toward less sustainable Games (Müller et al. 2021). Subsequent analysis of Tokyo 2020/21 documented improvements in environmental sustainability through a reduction in the number of constructed venues, reduced visitor footprint, and reduced number of attendees. Nevertheless, economic sustainability suffered because the share of private co-finance decreased, while the economic benefits are still unclear as they depend on the long-term viability of venues (Trendafilova et al. 2022).

Similarly, Leeds et al. (2022) found that economic benefits are often lower than expected because of substitution, crowding out, and leakage effects. Winner's curse and all-or-nothing demand curves are why cities decide to host mega events that make little economic sense, argue Little et al.

Contrary to these studies, Firgo (2021) established a significant boost for the regional GDP per capita for hosts of the Summer Olympics (but not the Winter Olympics) compared to applicant hosts who failed in their bids.

Research has also critically examined the upcoming Olympic Games, Paris 2024 and Los Angeles 2028.

Lauermann (2022) traces the urban planning history of Games hosted in the US. The study finds no American exceptionalism and shows no other host could repeat the supposedly successful LA 1984 model.[5] American hosts experienced cost overrun like everyone else, realised only weak economic gains, and saw politically damaging protests like Chicago's bid for the 2016 Olympics and Boston's bid for the 2024 Games. Lauermann found no support for the argument that the Olympics can achieve economic growth through public-private partnering and cost control. Instead, hidden cost overruns and public sector subsidies are the norm. Moreover, urban planning has changed from a primary focus on economic growth to include a broader set of policy goals.

---

[4] It should be noted that, for the first time, Tokyo 2020 saw a delay in the Games due to the Covid-19 pandemic. Thus, the mean schedule overrun of the Olympic Games is no longer 0% but 0.34%.

In sum, previous academic research on cost and cost overrun for the Olympic Games shows:

1. The issue of cost and cost overrun with the Games is now firmly accepted, which was not the case earlier;

2. Issues of cost and cost overrun have bearing on the sustainability of the Games;

3. The lack of sustainability of its flagship events is a critical, and acknowledged, challenge in keeping the Olympic Movement alive;

4. Attempts to reform the Games are underway but their impact remains to be seen.

---

[5] "Supposedly" because data have never been made available that documents that Los Angeles 1984 was in fact successful in terms of cost and cost overrun.

# Appendix B: Measuring Cost and Cost Overrun

Our method for measuring cost overrun is documented in Flyvbjerg and Stewart (2012) and Flyvbjerg et al. (2016, 2021). In short, we collected data on

1. *Operational costs* incurred by the Organizing Committee of the Olympic Games (OCOG) for "staging" the Games. The most significant components of this budget are technology, transportation, workforce, and administration, while other costs include security, catering, ceremonies, and medical services. These may be considered the variable costs of staging the Games and are formally called "OCOG costs" by the IOC, and
2. *Direct capital costs* incurred by the host city or country or private investors to build the competition venues, Olympic village(s), and international media and broadcast centre required to host the Games. These are the direct capital costs of hosting the Games and are formally called "non-OCOG direct costs."

Our data exclude *indirect capital costs* such as road, rail, airport infrastructure, hotel upgrades, or other business investments incurred in preparation for the Games but not directly related to staging the Games. These are wider capital costs, they are formally called "non-OCOG indirect costs," and typically they are substantial.

We sourced estimated costs from the candidature files. We collected actual costs from the final accounts published after the Games, which are accessible at the Olympic World Library (Olympics Study Centre n.d.).

Two exceptions are noteworthy:

1. Rio 2016 has yet to publish its final accounts. The latest estimates for actual cost were taken from the OCOG's published budgets and from accounts at relevant federal and local agencies.
2. For Paris 2024, we used the estimated budget of EUR 8.8 billion, as reported in January 2024 (Menocal Pareja 2024), when the preparations were more than 80% complete (Munana 2024). It should be emphasized again that judged by history there is a real risk that final outturn cost for Paris 2024 could end up higher than this estimate.

Milano-Cortina 2026, Los Angeles 2028, French Alps 2030, Brisbane 2032, and Salt Lake City-Utah 2032 are still preparing for the Games. We collected the latest budgets from the organizers' announcements. Los Angeles 2028 did not disclose a comprehensive budget in the candidature files, but cost estimates were sourced from public announcements at the launch of their bid. French Alps 2030 and Salt Lake City-Utah 2032 have yet to be officially selected, as such their candidature files are not public yet. Again, we collected data from the announcements made at the bid submission. It should be noted that French Alps 2030 is particularly vague with regard to venue costs, estimated between USD 0.7 and 1.4 billion.

We collected data for all Olympics between 1960 and 2024 (n=34). For 25 Games, data on estimated costs were available, and for 31 Games, data on actual costs were available.

To calculate cost overrun, we first removed inflation from actual cost data, which is reported in nominal terms, using GDP Deflators (World Bank, 2023a). Then, we calculated overrun as actual divided by estimated cost in local currencies in the constant price levels used in the respective candidature files.

For the cost analysis, we converted all local currencies into USD using average annual exchange rates provided by the World Bank (2023b). For the future Olympics, we used the latest 2023 average annual exchange rates to the USD, the IMF's GDP-deflation predictions for 2023 and 2024 for Paris (IMF 2023), and the long-range GDP deflator forecast by the OECD (2024) for Los Angeles 2028 and Brisbane 2032.

The method of treating the data is consistent with previous results in Flyvbjerg and Stewart (2012) and Flyvbjerg et al. (2016, 2021). The underlying GDP deflators and exchange rates are the only updates.

# References

BBC, 2016. *Rome 2024 Olympic Bid Collapses in Acrimony*, BBC, available at: https://www.bbc.co.uk/news/world-europe-37432928

Flyvbjerg, B., 2006. From Nobel Prize to project management: Getting risks right. *Project Management Journal*, 37(3), pp.5-15.

Flyvbjerg, B., 2011. Over Budget, Over Time, Over and Over Again: Managing Major Projects, in Peter W. G. Morris, Jeffrey K. Pinto, and Jonas Söderlund, eds., *The Oxford Handbook of Project Management*, Oxford: Oxford University Press, pp. 321-344.

Flyvbjerg, B., 2021. Make megaprojects more modular. *Harvard Business Review*, pp.58-63.

Flyvbjerg, B., Budzier, A., and Lunn, D., 2021. Regression to the tail: Why the Olympics blow up. *Environment and Planning A: Economy and Space*, 53(2), pp.233-260.

Flyvbjerg, B. and Gardner, D. 2023, *How Big Things Get Done: The Surprising Factors that Determine the Fate of Any Project from Home Renovations to Space Exploration and Everything in Between*, Penguin Random House.

Flyvbjerg, B. and Stewart, A., 2012. *Olympic Proportions: Cost and Cost Overrun at the Olympics 1960-2012*.

Flyvbjerg, B., Stewart, A., and Budzier, A., 2016. *The Oxford Olympics Study 2016: Cost and cost overrun at the games*. arXiv preprint arXiv:1607.04484.

Firgo, M., 2021. The causal economic effects of Olympic Games on host regions. *Regional science and urban economics*, 88, p.103673.

International Monetary Fund (IMF), 2023. *World Economic Outlook*. Available at: https://www.imf.org/external/datamapper/datasets

International Olympic Committee (IOC), 2015. *Olympic Agenda 2020 and Beyond*. Available at: https://olympics.com/ioc/olympic-agenda-2020

International Olympic Committee (IOC), 2020. *Olympic Agenda 2020+5*. Available at: https://olympics.com/ioc/olympic-agenda-2020-plus-5

International Olympic Committee (IOC), 2024, *Building Less, Better and Usefully*. Available at: https://olympics.com/en/paris-2024/information/infrastructures

Kahneman, D., Olivier S., and Sunstein, C. R., 2021, *Noise: A Flaw in Human Judgment*, William Collins.

Kelly, C., Karp, P., and Ore, A., 2023. *Australia Commonwealth Games 2026: Victoria cancels event after costs blow out to $7bn*, The Guardian, available at: https://www.theguardian.com/australia-news/2023/jul/18/australia-commonwealth-games-2026-victoria-cancels-event-after-funding-shortfall

Kirk, E., 2024. *IOC president says talks of cancelling 2032 Brisbane Olympic Games 'fake news'*, Newscorp, available at https://www.news.com.au/sport/olympics/ioc-president-says-talks-of-cancelling-2032-brisbane-olympic-games-fake-news/news-story/551f3710e71033f235f3e91dff47aed7

Lauermann, J., 2019. The urban politics of mega-events: Grand promises meet local resistance. *Environment and Society*, 10(1), pp.48-62.

Lauermann, J., 2022. The declining appeal of mega-events in entrepreneurial cities: From Los Angeles 1984 to Los Angeles 2028. *Environment and Planning C: Politics and Space*, 40(6), pp.1203-1218.

Leeds, M.A., Von Allmen, P., and Matheson, V.A., 2022. *The Economics of Sports*. Routledge.

Livingstone, R., 2015. *Hamburg 2024 Olympic Bid Officials Blame External Influences On Referendum Defeat*, GamesBids.com, available at: https://gamesbids.com/eng/summer-olympic-bids/2024-olympic-bid-news/hamburg-2024-olympic-bid-officials-blame-external-influences-on-referendum-defeat/

Lopes dos Santos, G. and Delaplace, M., 2023. Olympic Agenda 2020 and Paris 2024: Driving Change or Rhetoric as Usual?. *Journal of Olympic Studies*, 4(2), pp.56-89.

Menocal Pareja, M., 2024. *Paris 2024 facts and figures six months out*, Inside the Games, available at: https://www.insidethegames.biz/articles/1143455/numbers-paris-2024-nine-budget-spectator

Messenger, A., 2024. *No new stadium to be built for 2032 Olympics as Queensland opposition leader reveals plan*, The Guardian, available at: https://www.theguardian.com/sport/2024/mar/21/brisbane-olympics-2032-no-new-stadium-victoria-park-suncorp-gabba

Müller, M., Gogishvili, D., and Wolfe, S.D., 2022b. The structural deficit of the Olympics and the World Cup: Comparing costs against revenues over time. *Environment and Planning A: Economy and Space*, 54(6), pp.1200-1218.

Müller, M., Wolfe, S. D., Gaffney, C., Gogishvili, D., Hug, M., and Leick, A., 2021. Evaluating the Sustainability of the Olympic Games. *Nature Sustainability* 4 (4): 340-348.

Müller, M., Wolfe, S.D., Gogishvili, D., Gaffney, C., Hug, M., and Leick, A., 2022a. The mega-events database: systematising the evidence on mega-event outcomes. *Leisure studies*, 41(3), pp.437-445.

Munana, G., 2024. *84% of Paris 2024 Olympics construction finished*, Inside the Games, available at: https://www.insidethegames.biz/articles/1143290/84-of-construction-work-for-paris-2024

Nicoliello, M., 2021. The new agenda 2020+ 5 and the future challenges for the Olympic movement. *Athens Journal of Sports*, 8(2), pp.121-140.

OECD, 2024. *Real GDP long-term forecast*. Available at: https://data.oecd.org/gdp/real-gdp-long-term-forecast.htm

Olympic Studies Centre, n.d., *Olympic World Library – Official Reports*, available at: https://library.olympics.com/default/official-reports.aspx?_lg=en-GB

Preuss, H., 2004. *The economics of staging the Olympics: a comparison of the Games*, 1972-2008. Edward Elgar Publishing.

Preuss, H., Andreff, W., and Weitzmann, M., 2019. *Cost and revenue overruns of the Olympic Games 2000–2018* (p. 184). Springer Nature.

Preuss, H. and Weitzmann, M., 2023. Changes of costs, expenditures, and revenues between bidding and staging the Olympic games from Sydney 2000 to Tokyo 2020. *Event Management*, 27(3), pp.455-476.

Teh, C. and Stonington, J., 2022. *Beijing says the cost of hosting the 2022 Winter Games is among the cheapest ever at $3.9 billion. But the real cost might be more than $38.5 billion, 10 times the reported amount.* Business Insider, available at: https://www.businessinsider.com/real-cost-of-beijing-games-10-times-chinas-reported-figure-2022-1

Trendafilova, S., J. Ross, W., Triantafyllidis, S., and Pelcher, J., 2023. Tokyo 2020 Olympics sustainability: An elusive concept or reality?. *International Review for the Sociology of Sport*, 58(3), pp.469-490.

Vigor, A., Mean, M., and Tims, C. eds., 2004. *After the gold rush: A sustainable Olympics for London*. IPPR.

Waldenberger, F., 2020. Number Games: The economic impact of Tokyo 2020. In *Japan Through the Lens of the Tokyo Olympics* (pp. 18-24). Routledge.

Weitzmann, M. and Preuss, H., 2023. Key factors for cost overruns at Olympic Games-establishment of a model. *International Journal of Sport Management and Marketing*, 23(6), pp.506-522.

Wharton, D., 2017. *Budapest to withdraw bid for 2024 Olympics, leaving L.A. and Paris as only contenders*, LA Times, available at: https://www.latimes.com/sports/sportsnow/la-sp-budapest-2024-olympics-withdraw-20170222-story.html

Wolfe, S.D., 2023. Building a better host city? Reforming and contesting the Olympics in Paris 2024. *Environment and Planning C: Politics and Space*, 41(2), pp.257-273.

World Bank, 2022. *World Development Report 2022*, Chapter 5, available at: https://www.worldbank.org/en/publication/wdr2022/brief/chapter-5-managing-sovereign-debt

World Bank, 2023a. *GDP Deflator*, available at: https://data.worldbank.org/indicator/NY.GDP.DEFL.ZS

World Bank, 2023b. *Official exchange rate*, available at: https://data.worldbank.org/indicator/PA.NUS.FCRF

Zimbalist, A., 2020. *Circus Maximus* (3rd Edition). The Brookings Institution.

Zimbalist, A., Solberg, H.A., and Storm, R.K., 2024. The Olympic economy on the edge: can the Olympic Games survive its current financial model?. In *Research Handbook on Major Sporting Events* (pp. 693-708). Edward Elgar Publishing.